\documentclass[10pt]{article}
\usepackage[reqno]{amsmath}
\usepackage{amssymb,amsfonts,amsthm,upref,graphics,graphicx}

\newcommand{\R}{{\mathbb R}}

\newcommand{\N}{{\mathbb N}}
\newcommand{\Oo}{{\cal O}}
\newcommand{\gil}{[ \! [}
\newcommand{\gir}{] \! ]}
\newcommand{\betafcn}{{\mathbf{B}}}

\newtheorem{theo}{{\bf Theorem}}[section]

\newtheorem{lem}[theo]{{\bf Lemma}}

\numberwithin{equation}{section}

\author{Alfonso Agnew and Alain Bourget\thanks{\textbf{Mailing address}: Department of Mathematics, California State University (Fullerton),
McCarthy Hall 154, Fullerton CA 92834 (US).} \\ Department of Mathematics \\ California State University,
Fullerton }

\title{Semiclassical Density of States for the Quantum Asymmetric Top}

\date{}

\begin{document}

\maketitle

\begin{abstract}
In the quantization of a rotating rigid body, a {\it top,} one is concerned with the Hamiltonian operator $L_\alpha=\alpha_0^2 L_x^2 + \alpha_1^2 L_y^2 + \alpha_2^2 L_z^2,$ where $\alpha_0 < \alpha_1 <\alpha_2.$ An explicit formula is known for the eigenvalues of $L_\alpha$ in the case of the spherical top ($\alpha_1 = \alpha_2 = \alpha_3$) and symmetrical top ($\alpha_1 = \alpha_2 \neq \alpha_3$) \cite{LL}.  However, for the asymmetrical top, no such explicit expression exists, and the study of the spectrum is much more complex.
In this paper, we compute the semiclassical density of states for the eigenvalues of the family of operators $L_\alpha=\alpha_0^2 L_x^2 + \alpha_1^2 L_y^2 + \alpha_2^2 L_z^2$ for any $\alpha_0 < \alpha_1 <\alpha_2$.
\end{abstract}

\addtolength{\baselineskip}{1pt}

\section{Introduction}

Let $S^2 \subseteq \R^3$ be the 2-sphere and let $-\Delta_{S^2}$ be the constant curvature spherical
Laplacian on $S^2$. It is well known that the spectrum of $-\Delta_{S^2}$ consists of eigenvalues $\lambda$
given by
\begin{equation*}
  \lambda_k=  k(k+1), \ k=0,1,2,....
\end{equation*}
Moreover, the eigenspace corresponding to $\lambda_k$ is of dimension $2k+1$ and a basis of eigenfunctions is
obtained by taking the standard spherical harmonics of degree $k$, i.e.
$$Y_k^m(\theta,\phi)=P_k^m(\cos \theta) e^{im\phi}, \ |m| \leq k,$$ where $P_k^m$ is the associated Legendre
function of the first kind. For a more detailed treatment of the spectral theory of $\Delta_{S^2}$, we refer
the reader to \cite{Fo}.

From the fact that the eigenvalues $\lambda_k=k(k+1)$ of $-\Delta_{S^2}$ are of multiplicity $2k+1$, it is
easy to see that spectrum of $-\Delta_{S^2}$ has clustering. A nice way to illustrate this fact is to observe that 
for any Schwartz function $\varphi$ on $\R,$
\begin{equation} \label{sphere}
   \frac{1}{2k+1} \sum_{j=-k}^k \varphi\left( \frac{\sqrt{\lambda_k}}{k} \right) = \phi(1)+\Oo \left( \frac{1}{k}
   \right),
\end{equation}
in the semi-classical limit $k \to \infty$ (see \cite{M}). Expressions like those appearing on the RHS of
\eqref{sphere} are often referred to as a density of states (DOS) (see e.g.\cite{T1}). Together with the mean
level spacings and the pairs correlation, the DOS represents a useful quantity to measure the spread of the
spectrum.

In this paper we are interested in computing the DOS for $\sqrt{-L_\alpha},$ associated with the quantum asymmetric top with Hamiltonian $L_\alpha,$ where $L_\alpha$ is given by
\begin{equation*}
   L_\alpha := (\alpha_0^2 L_x^2 + \alpha_1^2 L_y^2 + \alpha_2^2 L_z^2),
\end{equation*}
and where
\begin{align*}
 L_{x}  & = -i(y \partial_z - z \partial_y),\\
 L_{y}  & =-i( z \partial_x - x \partial_z),\\
 L_{z}  & =-i(x \partial_y - y\partial_x).
\end{align*}
 Here, we assume that
$\alpha=(\alpha_0^2,\alpha_1^2,\alpha_2^2) \in \Lambda^3$, where
\begin{equation*}
   \Lambda^3:= \bigg\{ \alpha \in \R^3: 0<\alpha_0^2<\alpha_1^2<\alpha_2^2 \bigg\}
\end{equation*}
is the positive Weyl chamber.  It is well known that $-\Delta_{S^2}$ and $-L_\alpha$ are commuting,
self-adjoint, elliptic operators on $L^2(S^2)$ and therefore possess a Hilbert basis of joint eigenfunctions
-- the aforementioned spherical harmonics $Y_m^k$ \cite{BT}.  Moreover, it is easy to verify that their
principal symbols are linearly independent in $T^*(S^2)$. For these reasons, we say that $\Delta_{S^2}$ and
$L_\alpha$ form a quantum integrable system on $S^2$.  

An explicit formula is known for the eigenvalues of $L_\alpha$ in the case of the spherical top ($\alpha_1 = \alpha_2 = \alpha_3$) and symmetrical top ($\alpha_1 = \alpha_2 \neq \alpha_3$) \cite{LL}.  Although no such explicit formula exists for the eigenvalues of the asymmetrical top ($\alpha_1 \neq \alpha_2 \neq \alpha_3$), the spectrum was recently characterized in terms of parameters associated with the Lam\'e equation (cf.\ proposition 2.2 in \cite{T2}).

For such a system, it is customary to compute the DOS of their joint spectrum (see e.g.\ \cite{Ch,Co}). Here, we
are simply concerned with the density of states measures associated to the operators $\sqrt{-L_{\alpha}}$. In
the following, we denote by $E_k$ the eigenspace of $\Delta_{S^2}$ consisting of spherical harmonics of
degree $k$, i.e.\ $E_k=\text{Span}\{ Y_m^k : m=-k,-k+1,...,k \}$, and by $P_k$ the projection onto $E_k$. We
define the DOS measure associated to the operators $L^2_\alpha$ by
\begin{equation}
  d\rho_{DS}(x;k,\alpha):=\frac{1}{2k+1} \sum_{\lambda \in  \sigma(\sqrt{-P_k L_\alpha})}
  \delta \left(x-  \frac{\lambda}{k} \right)
\end{equation}
where $\sigma( \sqrt{-P_k L_\alpha})$ denotes the spectrum of $\sqrt{-P_k L_\alpha}$. Clearly, $\sigma(
\sqrt{-P_k L_\alpha})$ consists of the eigenvalues $\sqrt{\lambda_m^k}$, $m\leq |k|$, of $\sqrt{-L_\alpha}$
associated to the spherical harmonics of degree $k$. Our purpose here is to compute the density of states for
the measure $d\rho_{DS}(x,k;\alpha)$ in the semi-classical regime $k \to \infty$.

\subsection{Main result}

For any given $\alpha \in \Lambda^3$, let $g$ be the function defined on the rectangle $[0,\pi] \times
[0,\pi/2]$ by
\begin{equation*}
 g(\xi,\theta;\alpha)= (\alpha_1^2-\alpha_0^2)\left( \beta  \cos \xi + (\beta^2-1) \sin \theta
 \right)\sin \theta +\alpha_0^2.
\end{equation*}
where $\beta^2=\frac{\alpha_1^2-\alpha_0^2}{\alpha_2^2-\alpha_0^2}$. Finally, let
$g_+(\xi,\theta;\alpha)=\max\{0,g(\xi,\theta;\alpha)\}$.

\begin{theo} \label{main result}
Let $g_+$ be defined as above. Then,  we have that
\begin{equation*}
   \text{w-} \! \!  \lim\limits_{k \to \infty} d\rho_{DS}(x,k;\alpha) = \frac{1}{\pi} \int_{0}^{\pi}
   \int_{0}^{\pi/2} F(x;\theta,\xi,\alpha) \cos \theta \ d\xi d\theta
\end{equation*}
where $F$ is a convex combination of delta functions given by $$F(x;\theta,\xi,\alpha)= \frac{1}{4} \delta
\left( x- \frac{1}{2} \sqrt{g_+(\xi,\theta;\alpha)} \right) + \frac{3}{4} \delta \left( x- \frac{3}{2}
\sqrt{g_+(\xi,\theta;\alpha)} \right).$$ The weak limit is taken with respect to $C_c(\R^+)$.
\end{theo} \label{mainresult}

The proof of Theorem \ref{main result} is given in the third section of the paper. In the second section, we
show how one can separate the variables for the eigenvalue problem $-L_\alpha \psi = \lambda \psi$ and its
connection to the Lam\'e equation. In particular, we will show how the spectrum of the operators $-L_\alpha$
can be explicitly computed through the Lam\'e equation.

\section{Separation of variables and the Lam\'e equation}

As we mentioned earlier, $-\Delta_{S^2}$ and $-L_\alpha$ are commuting, self-adjoint, elliptic operators on
$L^2(S^2)$, hence they possess a Hilbert basis of joint eigenfunctions that form a class of spherical
harmonics.  Rather than working with the standard spherical harmonics $Y_k^m$, we introduce a more suitable
class of spherical harmonics for our purpose, the so-called \textit{Lam\'e harmonics} \cite{BT, WW}.

In terms of the Euclidean coordinates $(x,y,z) \in \R^3$, the Lam\'e harmonics of degree $k$ are written as
\begin{equation} \label{Lame harmonics1}
  \psi(x,y,z) = x^{\gamma_1} y^{\gamma_2} z^{\gamma_3} \prod_{j=0}^{\frac{1}{2}(k-|\gamma|)} \left(
  \frac{x^2}{\theta_j-\alpha_0^2}+\frac{y^2}{\theta_j-\alpha_1^2}+\frac{z^2}{\theta_j-\alpha_2^2} \right)
\end{equation}
where $\gamma_i \in \{0,1\}$ and $|\gamma|=\gamma_1+\gamma_2+\gamma_3$; the value of $|\gamma|$ is chosen so
that $k-\gamma$ is even. The values of the parameters $\theta_j$ are determined by the condition
$\Delta_{\R^3} \psi =0$. A simple computation shows that the $\theta_j$'s must satisfy Niven's equation
\begin{equation*}
  \sum_{j=0}^2 \frac{\gamma_j}{\theta_i-\alpha_j} + \sum_{j \neq i} \frac{1}{\theta_i-\theta_j} =0, \quad
  (i=1,...,\frac{1}{2}(k-|\gamma|)).
\end{equation*}

Based on Whittaker-Watson \cite{WW} terminology, we say that $\psi$ is of the first, second, third or fourth
species if $|\gamma|=0$, $|\gamma|=1$, $|\gamma|=2$ or $|\gamma|=3$ respectively. Note that there is no
Lam\'e harmonics of the second and fourth species for $k$ even, whereas for $k$ odd, there is none of the
first and third species. We will see later on that there exists respectively $k/2+1$, $3(k+1)/2$, $3k/2$ and
$(k-1)/2$ linearly independent Lam\'e harmonics of the first, second, third and fourth species. In
particular, for any positive integer $k$, there exist $2k+1$ linearly independent Lam\'e harmonics, hence
they form a basis for the space of spherical harmonics.

\subsection{Sphero-Conal coordinates}

In order to describe the Lam\'e harmonics in greater detail, it is useful to introduce a different system of
coordinates on $S^2$, namely the sphero-conal coordinates \cite{Sp, Vo}. We denote these by $(u_1,u_2)$. They
are defined for any given positive real constants $\alpha_0^2<\alpha_1^2<\alpha_2^2$ by the zeros of the
rational function
\[ R(u) = \frac{x^2}{u-\alpha_0^2} + \frac{y^2}{u-\alpha_1^2} + \frac{z^2}{u-\alpha_2^2}\]
where $(x,y,z) \in \R^3$. From the graph of $R(u)$, it is easy to see that $\alpha_0^2< u_1 < \alpha_1^2 <
u_2 < \alpha_2^2$.

\begin{figure}[htbp]
\begin{center}
\includegraphics[scale=0.5]{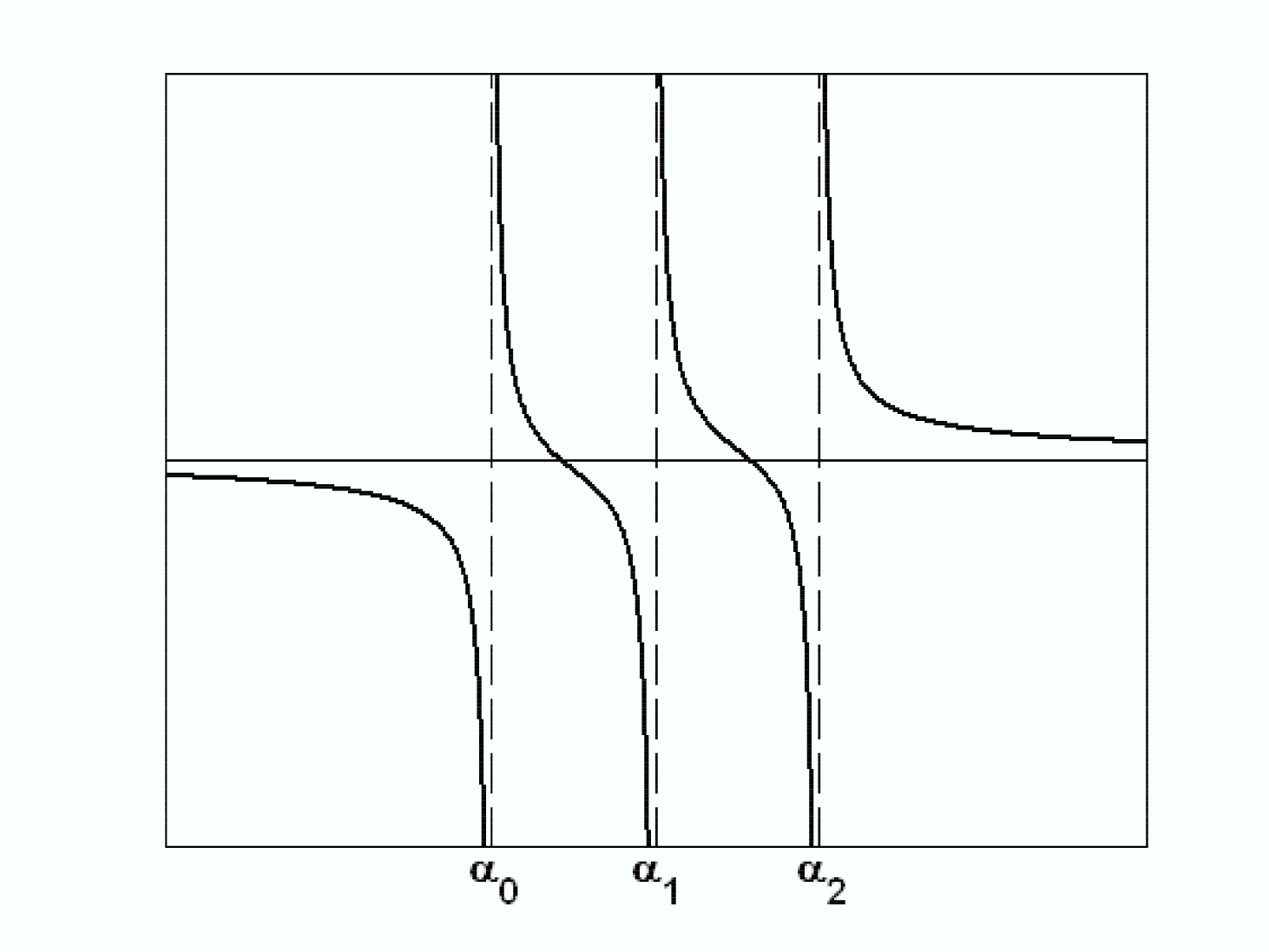}
\caption{ \small{The graph of $R(u)$ for fixed values of $x,y,z$ and $\alpha_i.$  The $\alpha_i$ correspond
to the vertical asymptotes. The intersections with the $u$-axis are the two roots of $R(u)$ corresponding to
the values of $u_i.$}}
\label{spheroconal}
\end{center}
\end{figure}

The equation $R(u)=0$ is invariant under rescaling $(x,y,z) \mapsto (tx,ty,tz)$, so the coordinates $(u_1,u_2)$ are
indeed coordinates on $S^2$ under the assumption $x^2+y^2+z^2=1$. They take their name from the fact that
they can be obtained by the intersection of the unit sphere with confocal cones.

The relations between the sphero-conal and Euclidean coordinates are given by
\begin{align*} \label{sphero-conal}
  x^2= & \frac{(u_1-\alpha_0^2)(u_2-\alpha_0^2)}{(\alpha_2^2-\alpha_0^2)(\alpha_1^2-\alpha_0^2)}, \\
  y^2= & \frac{(u_1-\alpha_1^2)(u_2-\alpha_1^2)}{(\alpha_2^2-\alpha_1^2)(\alpha_0^2-\alpha_1^2)}, \\
  z^2= & \frac{(u_1-\alpha_2^2)(u_2-\alpha_2^2)}{(\alpha_0^2-\alpha_2^2)(\alpha_1^2-\alpha_2^2)}.
\end{align*}
In particular, $(u_1,u_2)$ form an orthogonal system of coordinates on $S^2$. This can easily be seen by
considering the vectors $ \vec{r}_i=(\partial_{u_i} x, \partial_{u_i} y, \partial_{u_i} z)$ for which
\begin{eqnarray*}
 \vec{r}_1 \cdot \vec{r}_2 & = & \frac{x^2}{(u_1-\alpha_0^2)(u_2-\alpha_0^2)}+\frac{y^2}{(u_1-\alpha_1^2)(u_2-\alpha_1^2)}
 +\frac{z^2}{(u_1-\alpha_2^2)(u_2-\alpha_2^2)} \\
 &  = & \frac{R(u_1)-R(u_2)}{u_2-u_1} \\
 &  = & 0.
\end{eqnarray*}

\subsection{Separation of variables}

The great advantage of sphero-conal coordinates over other coordinate systems on $S^2$ is that they allow us
to simultaneously separate variables in both of the spectral problems for $-\Delta_{S^2}$ and $-L_{\alpha}$
(see \cite{Sp}). For example, in these coordinates, the Laplace equation $-\Delta_{S^2} \psi = k(k+1) \psi$
takes the form
\begin{equation} \label{Laplace}
\frac{4}{u_2-u_1}\sum_{i=1}^2 (-1)^i \left[ \sqrt{A(u_i)} \frac{\partial}{\partial u_i}\left( \sqrt{A(u_i)}
\frac{\partial \psi}{\partial u_i} \right) \right] =  k(k+1) \psi
\end{equation}
where $A(u_i)=(u_i-\alpha_0^2)(u_i-\alpha_1^2)(u_i-\alpha_2^2)$. One can then separate the variables and
write $\psi(u_1,u_2)=\psi_1(u_1)\psi_2(u_2)$. Denoting the separation constant by $-\lambda$, it follows
directly from \eqref{Laplace} that both $\psi_1$ and $\psi_2$ are solutions of the same \emph{Lam\'e
equation}
\begin{equation} \label{Lame}
  A(x) \psi_i''(x) + \frac{1}{2} A'(x) \psi_i'(x) =  \frac{1}{4}(k(k+1) x - \lambda) \psi_i(x) \qquad (i=1,2).
\end{equation}

From the general theory of Lam\'e equation \cite{WW}, it is well known that the solutions of \eqref{Lame} are
given by the Lam\'e functions
\begin{equation} \label{solpsi}
\psi_1(x)=\psi_2(x)=|x-\alpha_0^2|^{\gamma_1/2} |x-\alpha_1^2|^{\gamma_2/2} |x-\alpha_2^2|^{\gamma_3/2 }
\phi(x)
\end{equation}
where $\phi$ is a polynomial of degree $(k-|\gamma|)/2$ with $\gamma$ chosen as above. Consequently, the joint eigenfunctions of $-\Delta_{S^2}$
and $-L_\alpha$ are given by
\begin{equation} \label{Lame harmonics2}
  \psi(u_1,u_2) = \prod_{j=1}^2 |u_j-\alpha_0^2|^{\gamma_1/2} |u_j-\alpha_1^2|^{\gamma_2/2} |u_j-\alpha_2^2|^{\gamma_3/2 } \phi(u_j)
\end{equation}
Note that, up to a constant depending only on the $\alpha$'s and the solutions $\theta_j$'s of the Niven's equations, \eqref{Lame harmonics2} are the Lam\'e harmonics \eqref{Lame harmonics1} expressed in sphero-conal coordinates.

Based on these observations, we can now compute the eigenvalues of $-L_\alpha$. Let $E$ be such an
eigenvalue; we will show that $E=\lambda$, the separation constant obtained previously. First, we use the
fact that
\begin{equation*}
 -(L_x^2+L_y^2+L_z^2) \psi =-\Delta_{S^2} \psi  =  k(k+1) \psi
\end{equation*}
to deduce that
\[ ({\alpha_0^2-\alpha_1^2}) L_x^2 \psi + (\alpha_2^2-\alpha_1^2) L^2_z \psi = ( \alpha_1^2 k(k+1)-E) \psi . \]
In terms of sphero-conal coordinates, we can rewrite last equation as
\begin{multline*}
\frac{4}{u_2-u_1} \left[ (\alpha_1^2+u_2) \sqrt{A(u_1)} \frac{\partial}{\partial u_1} \left( \sqrt{A(u_1)}
\frac{\partial \psi}{\partial u_1} \right) \right. \\
\left. - (\alpha_1^2+u_1) \sqrt{A(u_2)} \frac{\partial}{\partial u_2} \left( \sqrt{A(u_2)} \frac{\partial
\psi}{\partial u_2} \right) \right] = (\alpha_1^2 k(k+1)-E) \psi.
\end{multline*}

Upon separating the variables, $\psi(u_1,u_2)=\psi_1(u_1)\psi_2(u_2)$, we obtain
\begin{equation} \label{Lame2}
  A(u_i) \psi_i''(u_i) + \frac{1}{2} A'(u_i) \psi_i'(u_i) = \frac{1}{4} (\mu u_i - E) \psi_i(u_i) \qquad (i=1,2).
\end{equation}

By comparison of \eqref{Lame2} with \eqref{Lame}, we conclude that $\mu=k(k+1)$ and $E=\lambda$ as desired.
All that remains to prove is that we get all the possible eigenvalues of $-L_{\alpha}$ in this way. This is a
consequence of the following result due to Stieltjes and Sz\"ego (see \cite{Sz}, \S 6.3):

\begin{theo} \label{Stieltjes-Szego}
Let $\rho_0,\rho_1,\rho_2$ be any three real positive numbers and let $a_1,a_2,a_3$ be any three real
distinct numbers. There exist exactly $m+1$ distinct real numbers $\nu$ for which the generalized Lam\'e
equation
\begin{equation} \label{genLame}
  A(x) y''(x) + \sum_{j=0}^2 \rho_j \prod_{i \neq j} (x-a_{i}) y'(x)
  = (m(m+1+|\rho|) x - \nu) y(x)
\end{equation}
has a polynomial solution $y$ of degree $m$. Moreover, the $m+1$ polynomial solutions obtained in this way
are linearly independent.
\end{theo}

Replacing the expression of the Lam\'e function $\psi_i$ given in \eqref{solpsi} into \eqref{Lame}, one can
easily verify that the polynomial $\phi$ of degree $(k-|\gamma|)/2$ satisfies the generalized Lam\'e equation
\begin{multline} \label{genLame2}
  A(x) \phi''(x) +  \sum_{j=0}^2 \left(\gamma_j+\frac{1}{2}\right) \prod_{l \neq j} (x-\alpha_l^2)
  \phi'(x) \\
  =  \frac{1}{4}\bigg((k-|\gamma|)(k+|\gamma|+1) x - \lambda + D(\alpha,\gamma) \bigg) \phi(x),
\end{multline}
where
$D(\alpha,\gamma)=(\alpha_0^2+\alpha_1^2)\gamma_2+(\alpha_0^2+\alpha_2^2)\gamma_1+(\alpha_1^2+\alpha_2^2)\gamma_0
+2\gamma_0\gamma_1\alpha_2^2 +2\gamma_1\gamma_2\alpha_0^2+2\gamma_0\gamma_2\alpha_1^2$. The values taken by
$\nu=\lambda-D(\alpha,\gamma)$ in terms of the different values of $\gamma$ are given in the table below.

\begin{table}[htdp]
\begin{center}
\begin{tabular}{|c|c|c|}
  \hline
  species & $\gamma_0,\gamma_1,\gamma_2$ & $\nu$ \\ \hline
  1 & $\gamma_0=\gamma_1=\gamma_2=0$    & $\lambda$ \\ \hline
    & $\gamma_0=1, \gamma_1=\gamma_2=0$ & $\lambda-\alpha_1^2-\alpha_2^2$ \\
  2 & $\gamma_1=1, \gamma_0=\gamma_2=0$ & $\lambda-\alpha_0^2-\alpha_2^2$ \\
    & $\gamma_2=1, \gamma_0=\gamma_1=0$ & $\lambda-\alpha_0^2-\alpha_1^2$\\ \hline
    & $\gamma_0=0, \gamma_1=\gamma_2=1$ & $\lambda-4\alpha_0^2-\alpha_1^2-\alpha_2^2$ \\
  3 & $\gamma_1=0, \gamma_0=\gamma_2=1$ & $\lambda-\alpha_0^2-4\alpha_1^2-\alpha_2^2$ \\
    & $\gamma_2=0, \gamma_0=\gamma_1=1$ & $\lambda-\alpha_0^2-\alpha_1^2-4\alpha_2^2$ \\ \hline
  4 & $\gamma_0=\gamma_1=\gamma_2=1$    & $\lambda-4(\alpha_0^2+ \alpha_1^2+\alpha_2^2)$ \\ \hline
\end{tabular}\end{center}
\label{table} \caption{The values taken by $\nu$}
\end{table}%

By Stieltjes' result with $\rho_i=\gamma_i+1/2$, we deduce that there are exactly $(k-|\gamma|)/2+1$ distinct
value $\nu$ for which \eqref{genLame2} has a polynomial solution $\phi$ of degree $(k-|\gamma|)/2$. In
particular, the number of Lam\'e harmonics of degree $k$ and of specie 1 is $k/2+1$, of species 2 is
$3(k+1)/2$, of specie 3 is $3k/2$ and of specie 4 is $(k-1)/2$. It follows that for any $k \in \N$, there
exist $2k+1$ linearly independent Lam\'e harmonics, so they form a Hilbert basis of $L^2(S^2)$.

Furthermore, for each $k \in \N$, we also obtain $2k+1$ values of $\nu$ (multiplicity included) to which
correspond by Table 1, $2k+1$ values of $\lambda$. In other words, the eigenvalues of the linearly
independent Lam\'e harmonics of degree $k$ are exactly given by the $2k+1$ values of $\lambda$. Therefore, we
have shown the first part of the following theorem.

\begin{theo}
Let $\alpha=(\alpha_0^2,\alpha_1^2,\alpha_2^2) \in \Lambda^3$, then the spectrum of the operator $-L_\alpha$
is given by all numbers $\lambda$ appearing on the RHS of the Lam\'e equation \eqref{Lame}. Moreover, the
$\lambda$'s corresponding to the Lam\'e harmonics of degree $k$ lie within the interval $(\alpha_0^2
(k-3)(k+1),\alpha_2^2 k(k+4)+ 4|\alpha|)$.
\end{theo}

The second part is an immediate consequence of a result due to Van Vleck \cite{Va} where he proves that all numbers
$\nu$ corresponding to the polynomial solutions of degree $m$ of the generalized Lam\'e equation
\eqref{genLame} lie inside the interval $(\alpha_0^2 m(m+1+|\rho|), \alpha_2^2 m(m+1+|\rho|))$. It follows
from this and \eqref{genLame2} that the eigenvalues $\lambda$ lie inside the interval
\[ \min_{\gamma} \{ \alpha_0^2 (k-|\gamma|)(k+|\gamma|+1) + D(\alpha,\gamma) \} \leq \lambda \leq \max_{\gamma} \{ \alpha_2^2
(k-|\gamma|)(k+|\gamma|+1) +D(\alpha,\gamma)\}. \]

Since $\gamma_i \in \{0,1\}$, it is then easy to see that
\[ \min_{\gamma} \{ \alpha_0^2 (k-|\gamma|)(k+|\gamma|+1) + D(\alpha,\gamma) \} \geq \alpha_0^2 (k-3)(k+1) \]
and
\[ \max_{\gamma} \{ \alpha_2^2 (k-|\gamma|)(k+|\gamma|+1) +D(\alpha,\gamma)\}  \leq \alpha_2^2 k(k+4) + 4|\alpha| \]
from which the conclusion of the theorem follows.

\section{Proof of Theorem 1.1}

Based on the different species of the eigenvalues, we partition the spectrum of $-L_\alpha$ into four disjoint
subsets $\sigma_1^k,...,\sigma_4^k$ defined by
\begin{equation*}
   \sigma_i^k:=\{ \lambda: \lambda \text{ is an eigenvalue of a Lam\'e harmonics of degree } k
   \text{ and of species } i \}.
\end{equation*}
For each $k \in \N$, we denote the eigenvalues of $\sqrt{-L_\alpha}$ corresponding to the $2k+1$ Lam\'e
harmonics of degree $k$ by
\begin{equation*}
    \sqrt{\lambda_{-k}^k(\alpha)} < \sqrt{\lambda_{-k+1}^k (\alpha)} < \cdots <  \sqrt{
   \lambda_k^{k}(\alpha)}.
\end{equation*}

Based on the definition of the $\sigma_i$, we can decompose $d\rho_{DS}(\varphi;k,\alpha)$ into four disjoints sums,
i.e.
\begin{eqnarray} \label{DS1}
  d\rho_{DS}(\varphi;k,\alpha) & = & \frac{1}{2k+1} \sum_{j=-k}^k \varphi \left( \frac{\sqrt{\lambda_j^k(\alpha)}}{k}
  \right) +\Oo\left( \frac{1}{k} \right) \nonumber\\
  & = & \frac{1}{2k+1} \sum_{i=1}^4 \sum_{\lambda \in \sigma_i^k} \varphi\left( \frac{\sqrt{\lambda}}{k} \right)
   + \Oo \left( \frac{1}{k} \right).
\end{eqnarray}

As we mentioned before, when $k$ is even, only the Lam\'e harmonics of the first and third species will
contribute to the sum above, whereas only the second and fourth species will contribute when $k$ is odd.
Therefore, we can write
\begin{equation*}
  \sum_{j=-k}^k \varphi\left(\frac{\sqrt{\lambda_j^k}}{k} \right) =
     \begin{cases}
          \sum_{\lambda \in \sigma_1^k} \varphi \left( \frac{ \sqrt{\lambda}}{k} \right) +
          \sum_{\lambda \in \sigma_3^k} \varphi \left( \frac{ \sqrt{\lambda}}{k} \right), & k \text{ even}\\
          \sum_{\lambda \in \sigma_2^k} \varphi \left( \frac{\sqrt{\lambda}}{k} \right) +
          \sum_{\lambda \in \sigma_4^k} \varphi \left( \frac{\sqrt{\lambda}}{k} \right), & k \text{ odd}.
     \end{cases}
\end{equation*}

The key observation here is that the eigenvalues can be obtained by simply regarding the polynomial solution
of the generalized Lam\'e equation \eqref{genLame2}. More precisely, we introduce the sets $Z_i$,
$i=1,2,3,4$, defined by
\[ Z_i^k := \{ \nu | \text{ There exist } \lambda \in \sigma_i^k \text{ and } \gamma \in \{0,1\}^3 \text{ such that }
\nu= \lambda - D(\alpha,\gamma) \}. \]

Consequently, the four sums above can now be taken over the sets $Z_i^k$ instead of $\sigma_i^k$. That is,
\begin{equation} \label{sums sigma z}
   \sum_{\lambda \in \sigma_i^k} \varphi\left( \frac{\sqrt{\lambda}}{k} \right) = \sum_{\nu \in Z_i^k} \varphi \left(
   \frac{\sqrt{\nu + D(\alpha,\gamma)}}{k} \right)
\end{equation}

Moreover, since $\varphi$ is compactly supported, we can approximate uniformly $\varphi$ by smooth functions.
Without loss of generality, we may therefore assume that $\varphi$ satisfies
\begin{equation*}
   \varphi\left( \frac{\sqrt{\nu + D(\alpha,\gamma)}}{k} \right) = \varphi \left( \frac{\sqrt{\nu}}{k} \right) +
    \Oo\left( \frac{1}{k} \right)
\end{equation*}
since $D(\alpha,\gamma)=\Oo(1)$. The equation \eqref{sums sigma z} easily implies that
\begin{equation} \label{df}
  \frac{1}{ |\sigma_i^k|} \sum_{\lambda \in \sigma_i^k} \varphi \left( \frac{ \sqrt{\lambda}}{k} \right) =
  \frac{1}{|Z_i^k|} \sum_{\nu \in Z_i^k} \varphi\left( \frac{\sqrt{\nu}}{k} \right)  + \Oo\left( \frac{1}{k}
  \right).
\end{equation}

The asymptotic of the sums in RHS of \eqref{df} are obtained through the following lemma.

\begin{lem} \label{mainlemma}
Let $\nu_0,...,\nu_m$ denote the $m+1$ real numbers for which the Lam\'e equation
\begin{equation*}
  A(x) y''(x) + \sum_{j=0}^2 \rho_j \prod_{i \neq j} (x-\alpha_i^2) y'(x) = (m(m+1+|\rho|) x - \nu) y(x)
\end{equation*}
admits a polynomial solution $y$ of degree $m$. For any $\varphi \in C_c(\mathbb{R}^+)$, we have
that
\begin{equation*}
  \frac{1}{m+1} \sum_{j=0}^m \varphi \left( \frac{\sqrt{\nu_j}}{m} \right) =\frac{1}{\pi} \int_{0}^{\pi}
  \int_{0}^{\pi/2} \varphi\left( \sqrt{g_+(\xi,\theta;\alpha)} \right)
   \cos \theta \ d\theta \ d\xi + \Oo\left( \frac{1}{m} \right)
\end{equation*}
where $g_+(\xi,\theta;\alpha)=  \max\{0,(\alpha_1^2-\alpha_0^2)\left( \beta \sin \theta \cos \xi + (\beta^2-1) \sin^2
\theta \right)+\alpha_0^2\}$, and $\beta^2=\frac{\alpha_2^2-\alpha_1^2}{\alpha_1^2-\alpha_0^2}$.
\end{lem}

The proof of Lemma \ref{mainlemma} is rather long and technical, so we prefer to postpone it until the end of
the present section. With this lemma in hand, we can now complete the proof of Theorem 1.1. As a consequence
of Lemma \ref{mainlemma}, we obtain for $k$ even,
\begin{eqnarray}\frac{1}{2k+1} \sum_{j=-k}^k \varphi\left( \frac{ \sqrt{\lambda_j^k(\alpha)}}{k} \right) 
 &  &\nonumber\\   &\hspace{-2cm}=&\hspace{-1cm}\frac{1}{2k+1} \left[ \sum_{\lambda \in \sigma_1^k}
     \varphi\left( \frac{\sqrt{\lambda}}{k} \right) + \sum_{\lambda \in \sigma_3^k}
     \varphi\left( \frac{\sqrt{\lambda}}{k} \right) \right]  \nonumber\\
    & \hspace{-2cm}=&\hspace{-1cm}\frac{1}{2k+1} \left[ \sum_{\nu \in Z_1^k} \varphi\left(\frac{\sqrt{\nu}}{k} \right) +
     \sum_{\nu \in Z_3^k} \varphi\left( \frac{\sqrt{\nu}}{k} \right) \right] +\Oo\left( \frac{1}{k} \right)\nonumber
     \end{eqnarray}
     \begin{eqnarray} \label{even}
   & =& \frac{1}{4} \left[ \frac{1}{k/2+1} \sum_{\nu \in Z_1^k} \varphi\left( \frac{1}{2} \frac{\sqrt{\nu}}{k/2}
     \right) \right] \nonumber\\
&+&  \frac{3}{4} \left[ \frac{1}{3k/2} \sum_{\nu \in Z_3^k} \varphi\left( \frac{3}{2} \frac{ \sqrt{\nu}}{3k/2}
     \right) \right] + \Oo\left( \frac{1}{k} \right)\end{eqnarray}

By Lemma \ref{mainlemma}, the first sum in the brackets of \eqref{even} is equal to
\begin{equation} \label{even2}
  \frac{1}{\pi} \int_{0}^{\pi}
  \int_{0}^{\pi/2} \varphi \left( \frac{1}{2} g_+(\xi,\theta;\alpha) \right)  \cos \theta \ d\theta \ d\xi + \Oo\left(
  \frac{1}{k} \right),
\end{equation}
and the second sum in brackets of \eqref{even} is equal to
\begin{equation} \label{even3}
  \frac{1}{\pi} \int_{0}^{\pi}
  \int_{0}^{\pi/2} \varphi \left( \frac{3}{2} g_+(\xi,\theta;\alpha) \right)  \cos \theta \ d\theta \ d\xi + \Oo\left(
  \frac{1}{k} \right),
\end{equation}

Combining equations \eqref{even2} and \eqref{even3}, we deduce that
\begin{equation} \label{even4}
 \begin{split}
  \frac{1}{2k+1} \sum_{j=-k}^k \varphi\left( \frac{ \sqrt{\lambda_j^k(\alpha)}}{k} \right) \\
  = \frac{1}{\pi} \int_0^{\pi} & \int_0^{\pi/2} F(\varphi;\xi,\theta;\alpha)
  \cos \theta \ d\theta \ d\xi + \Oo\left( \frac{1}{k} \right)
 \end{split}
\end{equation}
where the function $F$ is defined by $$F(\varphi;\xi,\theta;\alpha) =  \frac{1}{4} \varphi \left(
\frac{1}{2} g(\xi,\theta;\alpha) \right) + \frac{3}{4} \varphi \left( \frac{3}{2} g(\xi,\theta;\alpha)
\right).$$

The conclusion of Theorem 1.1 for $k$ even then follows from \eqref{DS1} and \eqref{even4}. Similarly, for
$k$ odd, we have that
\begin{eqnarray} \label{odd2}
 \lefteqn{ \hspace{-1cm}\frac{1}{2k+1} \sum_{j=-k}^k \varphi\left( \frac{ \sqrt{\lambda_j^k(\alpha)}}{k} \right)}\nonumber \\
     & = & \frac{1}{2k+1} \left[ \sum_{\lambda \in \sigma_2^k} \varphi\left( \frac{ \sqrt{\lambda}}{k}
     \right) + \sum_{\lambda \in \sigma_4^k} \varphi\left( \frac{\sqrt{\lambda}}{k}
     \right) \right] \nonumber\\
     & = & \frac{1}{2k+1} \left[ \sum_{\nu \in Z_2^k} \varphi\left(\frac{\sqrt{\nu}}{k} \right) +
       \sum_{\nu \in Z_4^k} \varphi\left( \frac{\sqrt{\nu}}{k} \right) \right] +\Oo\left( \frac{1}{k} \right) \nonumber\\
     & = & \frac{3}{4} \left[ \frac{1}{3k/2} \sum_{\nu \in Z_2^k} \varphi\left( \frac{3}{2} \frac{\sqrt{\nu}}{3k/2} \right)
         \right] \nonumber\\
     &   &  + \frac{1}{4} \left[ \frac{1}{(k-1)/2} \sum_{\nu \in Z_4^k} \varphi\left(\frac{1}{2} \frac{\sqrt{\nu}}{k/2}
     \right) \right] +\Oo\left( \frac{1}{k} \right).
\end{eqnarray}

As for the case $k$ even, we apply Lemma \ref{mainlemma} to conclude that \eqref{even4} holds when $k$ is a
positive odd integer.

To complete the proof of Theorem 1.1, it remains to prove Lemma \ref{mainlemma}.

\subsection{Proof of Lemma \ref{mainlemma}}

According to Theorem \ref{Stieltjes-Szego} with $a_0=-1$, $a_0=0$ and $a_2=\beta^2>0$, there exist $m+1$ real
values $\tilde{\nu_0},...,\tilde{\nu_m}$ for which the generalized Lam\'e equation
\begin{multline} \label{canonical}
x(x-\beta^2)(x+1) Y''(x)+ \left[  \rho_0 x(x-\beta^2) +  \rho_1 (x+1)(x-\beta^2) + \right.\\
\left.\rho_2x(x+1) \right] Y'(x) = (m(m+1+|\rho|) x - \tilde{\nu}) Y(x),
\end{multline}
admits a polynomial solution $Y$ of degree $m$. First, we show that for any $\varphi \in C_c(\mathbb{R}^+)$
\begin{equation} \label{SpecialCase}
  \frac{1}{m+1} \sum_{j=0}^m \varphi \left( \frac{\tilde{\nu}_j}{m^2} \right) =\frac{1}{\pi} \int_{0}^{\pi}
  \int_{0}^{\pi/2} \varphi\left( h(\xi,\theta;\alpha) \right) \cos \theta \ d\theta \ d\xi
  + \Oo\left( \frac{1}{m} \right)
\end{equation}
where $h(\xi,\theta;\alpha)= \beta \sin \theta \cos \xi + (\beta^2-1) \sin^2 \theta$.

The starting point in proving \eqref{SpecialCase} consists of establishing a three-term recurrence relation
satisfied by the Lam\'e polynomials $Y$. In particular, this will allow us to obtain the eigenvalues of
$-L_{\alpha}$ as the those of some tridiagonal matrix.

More precisely, we consider a Lam\'e polynomial of degree $m$ of the form
\[ Y(x) = \sum_{j=0}^m a_j x^j. \]
If we replace the expression for $Y(x)$ into the Lam\'e equation \eqref{canonical}, we obtain the
following three-term recurrence relation:
\begin{equation} \label{recurrence}
   A_j(\rho,\beta) a_j + B_j(\rho,\beta) a_{j+1} + C_j(\rho,\beta)  a_{j-1} = \tilde{\nu} a_j
   \qquad (j=0,...,m)
\end{equation}
where $a_{-1}=0$, $a_{m+1}=0$, and
\begin{equation}
  \begin{cases}
     A_j(\rho,\beta)= (\beta^2-1)j(j-1+\rho_1)-\rho_2 j + \beta^2 \rho_0 j  ,\\
     B_j(\rho,\beta)=(j+1)(j+\rho_1) \beta^2\\
     C_j(\rho,\beta)=\mu -(j-1)(j-2+|\rho|).
  \end{cases}
\end{equation}

Note that, as a result of the above, $A_0=B_m=C_0.$  These relations are more conveniently expressed in
matrix form. Indeed, if we introduce the tridiagonal matrix $A=(a_{ij})$, $i,j=0,...,m$, given by
\begin{equation}
a_{ij} = \left\{ \begin{array}{ll}
                      \frac{B_i(\rho,\beta)}{\mu}  & \text{ if } i=j-1\\
                      \frac{A_i(\rho,\beta)}{\mu}  & \text{ if } i=j\\
                      \frac{C_i(\rho,\beta)}{\mu}  & \text{ if } i=j+1,
                  \end{array} \right.
\end{equation}
then the three-term recurrence relation \eqref{recurrence} implies that
\begin{equation*}
 AX= \frac{\tilde{\nu}}{\mu} X,
\end{equation*}
where $X=(a_0,a_1,...,a_m)^T$. Throughout the rest of the proof, we denote by
$\frac{\tilde{\nu}_0}{\mu},...,\frac{\tilde{\nu}_m}{\mu}$ the $m+1$ eigenvalues of $A$.
Note that the components of the eigenvectors $X$ are exactly the coefficients of the Lam\'e polynomials $Y$.

We will divide the rest of the proof into several lemmas. The first one consists of computing the trace of
the powers $A^n$ for any $n \in \N$.

\begin{lem} \label{Lemma1}
We have that
\begin{multline}
  \text{Tr}(A^n) = \sum_{j=0}^{\gil n/2 \gir} \binom{n}{j,j,n-2j} \\
  \times \sum_{i=1}^m \left( 1- \frac{i^2}{m^2} \right)^j \left( \frac{i^2}{m^2} \right)^{n-j}
  (\beta^2-1)^{n-2j} \beta^{2j} + \Oo(1)
\end{multline}
for any positive integer $n$. Here, $\gil n/2 \gir$ denotes the greatest integer less or equal to $n/2$.
\end{lem}

\noindent\textit{Proof of Lemma \ref{Lemma1}:} We decompose $A$ as a sum of three matrices, $A=L+D+U$, where
$D=\frac{1}{\mu}\text{diag}(0,A_1,...,A_m)$ and
\[ L=\frac{1}{\mu}\begin{pmatrix}
  0 & 0 & 0 & \cdots & 0 & 0\\
  C_1 & 0 & 0 & \cdots & 0 & 0\\
  0 & C_2 & 0 & \cdots & 0 & 0\\
  \vdots & \vdots & \vdots &     & \vdots & \vdots \\
  0      & 0  & 0    & \cdots & C_m & 0
\end{pmatrix},
\, U= \frac{1}{\mu}\begin{pmatrix}
  0      &    B_0 &     0   & 0       & \cdots & 0 \\
  0      & 0      &   B_1   & 0       & \cdots & 0 \\
  \vdots & \vdots & \vdots  &  \vdots &        & \vdots \\
  0 & 0 & 0 & 0 & \cdots & B_{m-1} \\
  0 & 0 & 0 & 0 & \cdots & 0
\end{pmatrix}. \]

The trace of $A^n$ is then given by the trace of $(L+D+U)^n$. When we expand last expression, the
non-commutativity of the matrices $L,D$ and $U$ implies that the trace of $A^n$ is the sum of $3^n$ terms of the
form
\[ M_1 M_2 \cdots M_n, \]
where $M_i=L, D,$ or $U$.  This is unmanageable in its full generality for arbitrary $n.$ However, we are
interested primarily in the asymptotic information contained in the trace, which allows us to make
significant simplifications.

First, we point out that our need to consider $A^n$ stems from the fact that we will use polynomials to
approximate the continuous function $\varphi$ in Lemma \ref{mainlemma}.  Thus, we need to extract asymptotic
information about $\text{Tr}(A^n)$ for fixed, but arbitrary $n.$  In our case, we will ultimately be taking a
limit $m\rightarrow \infty$ for fixed $n,$ and so in this limit, $n/m \rightarrow 0$.

Secondly, we exploit the fact that the terms $ M_1 M_2 \cdots M_n$ are products of matrices, each being lower
diagonal ($L$), diagonal ($D$), or upper diagonal ($U$).  This allows us to make definite statements about
the \textit{zero structure} of the matrix products, i.e., the entries that are necessarily zero in the matrix
product.  For example, multiplication on the left or right by a diagonal matrix preserves the zero structure:
$LD$ and $DL$ are both lower diagonal if $L$ is.  The analogous statement holds for $UD$ and $DU.$  The
effect of multiplying by $L$ or $U$ is only slightly less simple.  In fact, as far as the effect on zero
structure is concerned, $L$ and $U$ behave like quantum mechanical creation and annihilation operators
(respectively).  In detail, if we denote by $R_M$ (resp.\ $L_M$) the operation of right (resp.\ left)
multiplication by a matrix $M$, then for any matrix $B$:
\begin{enumerate}
\item[(i)] $R_UB$ corresponds to shifting all columns of B one place to the right:
$\mathrm{col}_{i+1}(R_UB) = \mathrm{col}_i(B)$, creating a zero column in the first column.
\item[(ii)] $R_LB$ corresponds to shifting all columns of B one place to the left: \- $\mathrm{col}_{i-1}(R_UB) = \mathrm{col}_i(B)$, creating a zero column in the last column.
\item[(iii)] $L_UB$ corresponds to shifting all rows of B up one place: $\mathrm{row}_{i-1}(R_UB) = \mathrm{row}_i(B)$, creating a zero row in the last row.
\item[(iv)] $L_UB$ corresponds to shifting all rows of B down one place: $\mathrm{row}_{i+1}(R_UB) = \mathrm{row}_i(B)$, creating a zero row in the first row.
\end{enumerate}
As a result, the diagonal of a term $M_1 M_2 \cdots M_n$ in $A^n$ will have zero trace unless the number of
factors $j$ of $L$ is the same as the number of factors of $U.$  The remaining $n-2j$ factors must all be
$D.$  Thus, many of the $3^n$ terms do not contribute to $\text{Tr}(A^n).$

The last issue concerns the lack of commutativity in the terms that do contribute to the trace.   Some of
these terms are of the form
\begin{equation}\label{canonicalterm}
(LU)^j D^{n-2j},\, j=0, \dots, \gil n/2 \gir.
\end{equation}
Since $LU$ and $D$ are diagonal, the trace is particularly simple to compute in the case of the
\textit{canonical terms} (\ref{canonicalterm}):
\begin{eqnarray*}
 \text{Tr}(M_1 M_2 \cdots M_n) & = & \text{Tr}((LU)^j D^{n-2j})  \\
      & = & \sum_{i=1}^m \left( 1- \frac{i^2}{m^2} \right)^j \left( \frac{i^2}{m^2} \right)^{n-j}
            (\beta^2-1)^{n-2j} \beta^{2j} + \Oo(1).
\end{eqnarray*}

Noncanonical terms will differ from canonical terms only at order $\Oo(n/m) = \Oo(1/m),$ and so for
asymptotic purposes, we may assume that all terms have the canonical form (\ref{canonicalterm}).  To see
this, note that multiplication of matrices of the form $L, D,$ and $U$ constitutes a shifting of their rows
and columns.  For terms with $n$ factors, the number of shifts is at most $n.$  Being products of matrices
that are (lower, upper) diagonal, the noncanonical terms will yield sums of products of the form
\[ \Gamma_p\Delta_{q}, \]
where $\Gamma_p, \Delta_{q} \in \{A_l/\mu, B_l/\mu, C_l/\mu\, | \, l = 0, 1, \dots, m\}$ and $|p-q| =
\Oo(n).$  As an example,
\begin{eqnarray*}
\frac{A_p}{\mu}\frac{B_{q}}{\mu} &=&  \beta^2 (\beta^2-1)\frac{p^2q^2}{m^4}  + \Oo\left( \frac{1}{m} \right)\\
&=& \beta^2 (\beta^2-1)\frac{p^2(p + \Oo(n))^2}{m^4} + \Oo\left( \frac{1}{m} \right)\\
&=& \beta^2 (\beta^2-1)\frac{p^4}{m^4} + \Oo(n/m) + \Oo\left( \frac{1}{m} \right)\\
&=& \beta^2 (\beta^2-1)\frac{p^4}{m^4} + \Oo\left( \frac{1}{m} \right).
\end{eqnarray*}

Since there are exactly $\binom{n}{j,j,n-2j}$ matrices $M_1 M_2 \cdots M_n$ that contain $j$ factors of $L$,
$j$ factors of $U$ and $(n-2j)$ factors of $D$, we finally deduce that

\begin{eqnarray}
  \text{Tr}(A^n) & = & \sum_{j=0}^{\gil n/2 \gir} \binom{n}{j,j,n-2j} \text{Tr}((LU)^j D^{n-2j})
                          + \Oo(1)  \nonumber \\
                    & = & \sum_{j=0}^{\gil n/2 \gir} \binom{n}{j,j,n-2j} \sum_{i=1}^m \left( 1- \frac{i^2}{m^2} \right)^j
                          \left( \frac{i^2}{m^2} \right)^{n-j} (\beta^2-1)^{n-2j} \beta^{2j} \nonumber \\
                    &   &       + \Oo(1). \label{trace A}\label{traceeq1}
\end{eqnarray}

This completes the proof of the Lemma. \qed\\

The next result deals with the inner sum $\sum_{i=1}^m \left( 1- \frac{i^2}{m^2} \right)^j \left( \frac{i^2}{m^2} \right)^{n-j}$ in (\ref{traceeq1}). As the next lemma shows, this sum is asymptotically given by a Beta integral.

\begin{lem} \label{Lemma3}
We have that
\begin{equation} \label{lem2eq1}
   \frac{1}{m} \sum_{i=0}^m \left( 1- \frac{i^2}{m^2} \right)^j \left( \frac{i^2}{m^2} \right)^{n-j}
   = \frac{1}{2} \betafcn\, (j+1,n-j+1/2) + \Oo \left( \frac{1}{m} \right)
\end{equation}
where $\betafcn(p,q)$ is the standard Beta integral defined by
\[ \betafcn(p,q)=2 \int_0^{\pi/2} \cos^{2p-1} \theta \ \sin^{2q-1} \theta \ d\theta. \]
\end{lem}

\noindent\textit{Proof of Lemma \ref{Lemma3}:} This is obvious. The LHS of \eqref{lem2eq1} is a Riemann sum
for the function $(1-x^2)^j (x^2)^{n-j}$ on $[0,1]$, hence
\begin{equation*}
 \frac{1}{m} \sum_{i=0}^m \left( 1- \frac{i^2}{m^2} \right)^j \left( \frac{i^2}{m^2} \right)^{n-j}=
    \int_0^1 (1-x^2)^j (x^2)^{n-j} \ dx + \Oo \left( \frac{1}{m} \right).
\end{equation*}
The conclusion of the lemma follows by making the substitution $x=\sin \theta$ and using the trigonometric representation
of the Beta integral. \qed \\

As a consequence of \eqref{trace A} and Lemma \ref{Lemma1}, it follows that
\begin{eqnarray}
  \frac{1}{m} \text{Tr}(A^n) & = & \frac{1}{m} \sum_{i=0}^{m} \left(\frac{\nu_i}{\mu}\right)^n \nonumber \\
    & = & \frac{1}{2} \sum_{j=0}^{\gil n/2 \gir} \binom{n}{j,j,n-2j} \betafcn \, (j+1,n-j+1/2)
          (\beta^2-1)^{n-2j} \beta^{2j} \nonumber \\
    &   & + \Oo\left( \frac{1}{m} \right). \label{sum over j}
\end{eqnarray}

In order to evaluate the sum inside the integral sign, we use the \textit{sinc} function defined by
\begin{equation*}
   \text{sinc}(x) = \begin{cases}
                        1 & \text{ for } x=0, \\
                        \frac{\sin x}{x} & \text{ for } x \neq 0.
                    \end{cases}
\end{equation*}
The key point here is to observe that $\text{sinc}(\pi x)=0$ when $x$ is a non-zero integer, and that
$\text{sinc}(0)=1$. Using this function, we can then replace the sum in \eqref{sum over j} by the more
appropriate sum over multi-index $\gamma=(\gamma_1,\gamma_2,\gamma_3)$ such that $|\gamma|=n$. More
precisely, we have
\begin{equation} \label{sum}
  \begin{split}
    \shoveleft{\sum_{j=0}^{\gil n/2 \gir} \binom{n}{j,j,n-2j} \betafcn \, (j+1,n-j+1/2) \, (\beta^2-1)^{n-2j}
    \beta^{2j} }\\
      = \sum_{|\gamma|=n} \binom{n}{\gamma} (\beta^2-1)^{\gamma_3} \beta^{\gamma_1+\gamma_2}
       \, \betafcn  \left( \gamma \right) \text{sinc}  (\pi(\gamma_1- & \gamma_2)).
   \end{split}
\end{equation}
where $\betafcn(\gamma):=\betafcn\left( \frac{\gamma_1}{2}+\frac{\gamma_2}{2}+1,
n-\frac{\gamma_1}{2}-\frac{\gamma_2}{2}+\frac{1}{2} \right)$.  Based on the representation of
$\text{sinc}(x)$ as the integral
\begin{equation}
\text{sinc}(\pi(\gamma_1-\gamma_2)) = \frac{1}{2\pi} \int_{-\pi}^{\pi} e^{i\xi(\gamma_1-\gamma_2)}  \ d \xi,
\end{equation}
the RHS of \eqref{sum} can be written as
\begin{equation} \label{sum2}
   \frac{1}{2\pi} \int_{-\pi}^{\pi} \sum_{|\gamma|=n} \binom{n}{\gamma} (\beta^2-1)^{\gamma_3}
   \beta^{\gamma_1+\gamma_2} \, \betafcn\left( \gamma \right)  e^{i\xi(\gamma_1-\gamma_2)} \ d\xi.
\end{equation}

Replacing $\betafcn(\gamma)$ by the expression
\begin{equation*}
  \betafcn(\gamma) = 2 \int_0^{\pi/2} (\cos \theta)^{\gamma_1+\gamma_2+1} (\sin
  \theta)^{2n-\gamma_1-\gamma_2} \ d\theta,
\end{equation*}
we can then use the Multinomial Theorem to evaluate the sum in \eqref{sum2}. We obtain
\begin{equation} \label{multinomial}
 \begin{split}
   \sum_{|\gamma|=n} \binom{n}{\gamma} (\beta^2-1)^{\gamma_3} \beta^{\gamma_1+\gamma_2}
   (\cos \theta)^{\gamma_1+\gamma_2} (\sin \theta)^{2n-\gamma_1-\gamma_2} e^{i\xi(\gamma_1-\gamma_2)}\\
   = (  \beta \cos \xi  \sin 2\theta + (\beta^2-1) \sin^2 \theta)^n.
 \end{split}
\end{equation}

If we denote by $h(\xi,\theta)= (\beta^2 \cos \xi  \sin 2\theta +(\beta^2-1) \sin^2 \theta)$, then equations
\eqref{sum over j} through \eqref{multinomial} imply that
\begin{equation*}
    \frac{1}{m} \text{Tr} (A^n) =
    \frac{1}{2\pi} \int_{-\pi}^{\pi} \int_0^{\pi/2} h^n(\xi,\theta)  \cos \theta \ d\theta \
    d\xi + \Oo \left( \frac{1}{m} \right).
\end{equation*}

The rest of proof of Lemma \ref{mainlemma} follows by the standard functional calculus on the Banach algebra
$M_m(\R)$, the set of all matrices of order $m$ with real entries. However, we can also complete the proof by
simply observing that for any polynomial $P$,
\begin{equation}\label{dblint}
    \frac{1}{m} \sum_{i=0}^{m} \mathrm{Tr}(P( A )) = \frac{1}{2\pi} \int_{-\pi}^{\pi} \int_0^{\pi/2}
    P(h(\xi,\theta))  \cos \theta \ d\theta \ d\xi + \Oo \left( \frac{1}{m} \right).
\end{equation}

Finally, Weierstrass' Theorem implies that for any compactly continuous function $\varphi$ and any $\epsilon >0,$ there exists a
polynomial $P$ with 
\begin{equation}\label{weier}
\sup_x |\varphi(x)-P(x)| < \epsilon/3.
\end{equation}

This implies
\begin{equation}\label{temp1}
\frac{1}{2\pi}\int_{-\pi}^\pi \int_0^{\pi/2}|\phi(h(\xi,\theta))-P(h(\xi,\theta))|\cos \theta\ d\theta\ d\xi < \epsilon/3.
\end{equation}

From the Spectral Mapping Theorem and \eqref{weier} we obtain
\begin{equation}
\left| \frac{1}{k}\mathrm{Tr}(\varphi(A)) - \frac{1}{k}\mathrm{Tr}(P(A)) \right| < \epsilon/3.
\end{equation}

We choose $m$ big enough in \eqref{dblint} so that 
\begin{equation}\label{temp2}
\left| \frac{1}{m} \sum_{i=0}^{m} \mathrm{Tr}(P( A )) - \frac{1}{2\pi} \int_{-\pi}^{\pi} \int_0^{\pi/2}
    P(h(\xi,\theta))  \cos \theta \ d\theta \ d\xi \right| <\epsilon/3.
\end{equation}

As a consequence of \eqref{temp1} - \eqref{temp2}
\begin{equation} \label{trace4}
  \lim_{m\to \infty}  \frac{1}{m} \sum_{i=0}^{m} \mathrm{Tr}(\varphi\left( A \right)) = \frac{1}{2\pi} \int_{-\pi}^{\pi}
    \int_0^{\pi/2} \varphi(h(\xi,\theta))  \cos \theta \ d\theta \ d\xi.
\end{equation}

This completes the proof of \eqref{SpecialCase} for $a_0=-1$, $a_1=0$ and $a_2=\beta^2$. In the case of interest
to us, namely $a_0=\alpha_0^2$,  $a_1=\alpha_1^2$ and $a_2=\alpha_2^2$, we use the fact that the
Lam\'e equation is invariant under affine transformations to make the change of variable $x \mapsto
x(\alpha_1^2-\alpha_0^2)+\alpha_0^2$. If we let $y$ be the function defined by
$y(x):=Y(x(\alpha_1^2-\alpha_0^2)+\alpha_1^2)$, then it is not hard to show that $y$ satisfies the standard Lam\'e equation
\begin{equation*}
  A(x) y''(x) + \sum_{j=0}^2 \rho_j \prod_{i \neq j} (x-\alpha_i^2) y'(x) = (\mu x - \nu) y(x)
\end{equation*}
where $\nu=\tilde{\nu}(\alpha_1^2-\alpha_0^2)+\alpha_1^2\mu$. From the fact that $-1\leq
\frac{\tilde{\nu}}{\mu} \leq \beta^2$, we easily deduce that $\alpha_0^2 \leq \frac{\nu}{\mu} \leq
\alpha_2^2$.

Furthermore, if we introduce the function $\varphi_{\alpha}(x):=\varphi \left(
x(\alpha_1^2-\alpha_0^2)+\alpha_1^2 \right)$ for any $\varphi \in C_c(\R^+)$, then we obtain that
\begin{eqnarray}
   \frac{1}{m} \sum_{i=0}^m \varphi \left( \frac{\nu_i}{m^2} \right) & = & \frac{1}{m} \sum_{i=0}^m
               \varphi \left( \frac{\tilde{\nu}_i}{m^2} (\alpha_1^2-\alpha_0^2)+\alpha_1^2 \right)
               + \Oo \left( \frac{1}{m} \right)\nonumber \\
        & = & \frac{1}{m} \sum_{i=0}^m \varphi_\alpha \left( \frac{\tilde{\nu}_i}{m^2} \right)
              + \Oo \left( \frac{1}{m} \right). \label{last}
\end{eqnarray}
It then follows by \eqref{trace4} and \eqref{last} that
\begin{equation}
 \lim_{m\to \infty} \frac{1}{m} \sum_{i=0}^m \varphi \left( \frac{\nu_i}{m^2} \right) = \frac{1}{2\pi} \int_{-\pi}^{\pi}
    \int_0^{\pi/2} \varphi(g(\xi,\theta;\alpha))  \cos \theta \ d\theta \ d\xi
\end{equation}
where $g(\xi,\theta;\alpha)=h(\xi,\theta;\alpha)(\alpha_1^2-\alpha_0^2)+\alpha_1^2$.
Since $\varphi$ is supported in $\R^+$, last equation remains valid if we replace $g(\xi,\theta;\alpha)$ by
$g_+(\xi,\theta;\alpha)=\max\{0,g(\xi,\theta;\alpha)\}$ and $\varphi(x)$ by $\varphi(\sqrt{x})$. This
completes the proof of Lemma 3.1. \qed

\end{document}